%% file: MS2510.tex
\documentstyle[psfig]{aa}
\def\lsim{~\rlap{$<$}{\lower 1.0ex\hbox{$\sim$}}}  
\def\gsim{~\rlap{$>$}{\lower 1.0ex\hbox{$\sim$}}} 

\begin{document}

	\title{Unabsorbed Seyfert 2 galaxies}

        \author{F.~Panessa\inst{1,2} and L.~Bassani\inst{1}}
    	\offprints{F. Panessa (panessa@bo.iasf.cnr.it)}

    	\institute{IASF/CNR, via Piero Gobetti 101, I-40129 Bologna, 
              Italy \and 
              Dipartimento di Astronomia, Universita' di Bologna,
	      via Ranzani 1, I-40127 Bologna, Italy}

	\date{Received / accepted}

	\titlerunning{Unabsorbed Seyfert 2 galaxies}
	\authorrunning{F.~Panessa and L.~Bassani}

\abstract{
We present a sample of 17 type 2 Seyfert galaxies
which have an X-ray column density lower than 10$^{22}$ cm$^{-2}$.
The Compton thin nature of these sources is strongly suggested by 
isotropic indicators. We estimate the fraction of these sources
to be in the range of 10\% - 30\% of the population of
type 2 Seyfert galaxies. Furthermore,
this fraction appears to increase progressively at lower luminosities.
The simple formulation of the Unified Model for Seyfert galaxies
is not applicable in such sources since the pc-scale molecular torus is not 
likely to be responsible for the low column density observed, 
instead the absorption observed is likely to originate at larger scales. 
According to this hypothesis, in these objects the broad line 
regions are covered by some dusty obscuring material. 
In particular, this could occur in objects
with dust lanes, patches or HII regions. However, we cannot rule out that
in the lowest luminosity sources the BLR is weak, absent or has faded away. 
This last scenario is consistent with the predictions of some recent 
theoretical models for low luminosity AGNs.
}

    \maketitle

\section{Introduction}

Seyfert galaxies belong to the class of Active Galactic Nuclei (AGN). 
According to the standard model, in active galaxies an accretion disk
around a massive black hole produces a hard X-ray continuum, which 
photoionizes the Broad Line Region (BLR, where broad emission lines originate) 
and the Narrow Line Region (NLR, where narrow emission lines originate) 
located at $<$ 1 pc and at $<$ 100 pc from the nuclear engine respectively.
Seyfert galaxies are classified as type 1 or type 2.
Type 1 have both narrow forbidden lines 
(FWHM $\leq$ 10$^{3}$ km/s) and broad Balmer lines 
(FWHM $\sim$ 10$^{4}$ km/s) in their optical spectrum, while type 2 have 
only narrow lines. Actually they are the same object: 
type 2 Seyferts harbour a BLR, but this is obscured from view in 
some directions by a molecular torus
(Unification Model; Antonucci, 1993).

Optical spectropolarimetry measurements of scattered broad permitted lines
provide strong evidence in favour of the unified model 
(Antonucci \& Miller 1985). At least 35\% of Seyfert 2 galaxies have 
broad emission lines seen in polarized light 
(Tran 2001; Moran et al. 2000), therefore a good fraction of Seyfert 2 
galaxies seem to host a hidden Seyfert 1 nucleus.

More evidence in favour of the unified model comes from
the X-ray spectra: the column density  of neutral hydrogen 
in type 2 Seyferts is significantly higher than in type 1 objects 
as  would be expected if, for the type 2 sources,
the nucleus is observed through the torus 
(Turner et al. 1997; Smith \& Done 1996).
Observed column  densities range from 10$^{22}$ cm$^{-2}$ to higher than
10$^{24}$ cm$^{-2}$ for $\sim$ 96\% of the objects (Risaliti et al. 1999;
Bassani et al. 1999). 

However, not all Seyfert 2 galaxies have a Broad Line Region in polarized 
light and not all Seyfert 2 galaxies have column densities higher than 
10$^{22}$ cm$^{-2}$. Polarimetric surveys of complete samples of Seyfert 2s
indicate that a large fraction of these objects (up to 50\%) do not show 
a hidden BLR typical of an obscured Seyfert 1 nucleus. Furthermore
there have been some recent examples of Seyfert 2 galaxies, 
such as NGC 3147 (Ptak et al. 1996), NGC 4698 
(Pappa et al. 2001) and NGC 7590 (Bassani et al. 1999), which have no or 
low absorption measured from the X-ray spectrum. It can be argued that these 
are Compton thick objects i.e. in which the medium is
thick to Compton scattering such that the transmitted component is 
completely suppressed below 10 keV and the 2-10 keV spectrum is 
dominated by reprocessed components. In this case the hard X-ray spectrum
is characterized by a flat Compton reflection component from the inner
surface of the torus and/or a steeper component ascribed to an ionized,
warm scattering medium. When the absorbing medium has column density 
N$_{H}$ $>$ 10$^{24}$ cm$^{-2}$, then the transmitted component
can be observed above 10 keV.
Therefore, in these sources the true column density
can only be estimated by higher energy data 
for N$_{H}$ $>$ 10$^{24}$ cm$^{-2}$
or measured indirectly by comparing the X-ray luminosity with the
Far-Infrared or [OIII] luminosities for even higher column densities.
In the above mentioned sources these absorption indicators  
suggest that they are actually Compton thin objects.
At the moment at least $\sim$ 4\% of Seyfert 2s have N$_{H}$ $<$ 
10$^{22}$ cm$^{-2}$ (Risaliti et al. 1999). The exact nature of
these peculiar Seyfert 2s is still unclear, as it is not obvious what
obscures their Broad Line Region: they may be intrinsically different
objects than those explained by the unified theory or in other words
they may be the ``true'' Seyfert 2 galaxies which are sometimes discussed 
in the literature (Tran 2001).  

In this paper we have collected a sample of Seyfert 2 galaxies 
characterized by low X-ray absorption in order to study their properties, 
estimate their abundance and understand better their nature.

\section{The sample}

The sample consists of objects which are classified in NED 
as type 2 Seyferts\footnote{In type 2 optical classification we also include 
Seyferts 1.9 and 1.8.} and are characterized by low absorption in X-rays 
(N$_{H}$ $\lsim$ 10$^{22}$ cm$^{-2}$).

\input{MS2510tab1.tex}

The 17 objects of the sample 
are listed in table \ref{tab:tab_campione} with optical
positions in equatorial coordinates for epoch J2000, redshift {\it z}
as reported in NED, Galactic column density from 21 cm measurement 
in units of 10$^{20}$ cm$^{-2}$ obtained from the HEASARC (High Energy
Astrophysics Science Archive Research Center) on-line service and
NED classification (S=Seyfert, L=Liner, SB=Starburst). 
In table \ref{tab:tab_x} we list the main X-ray spectral parameters 
in addition to the Infrared and [OIII]$\lambda$5007 flux values.
    
\subsection{Optical classification}

\begin{figure}[!htbp]
\begin{center}
\psfig{figure=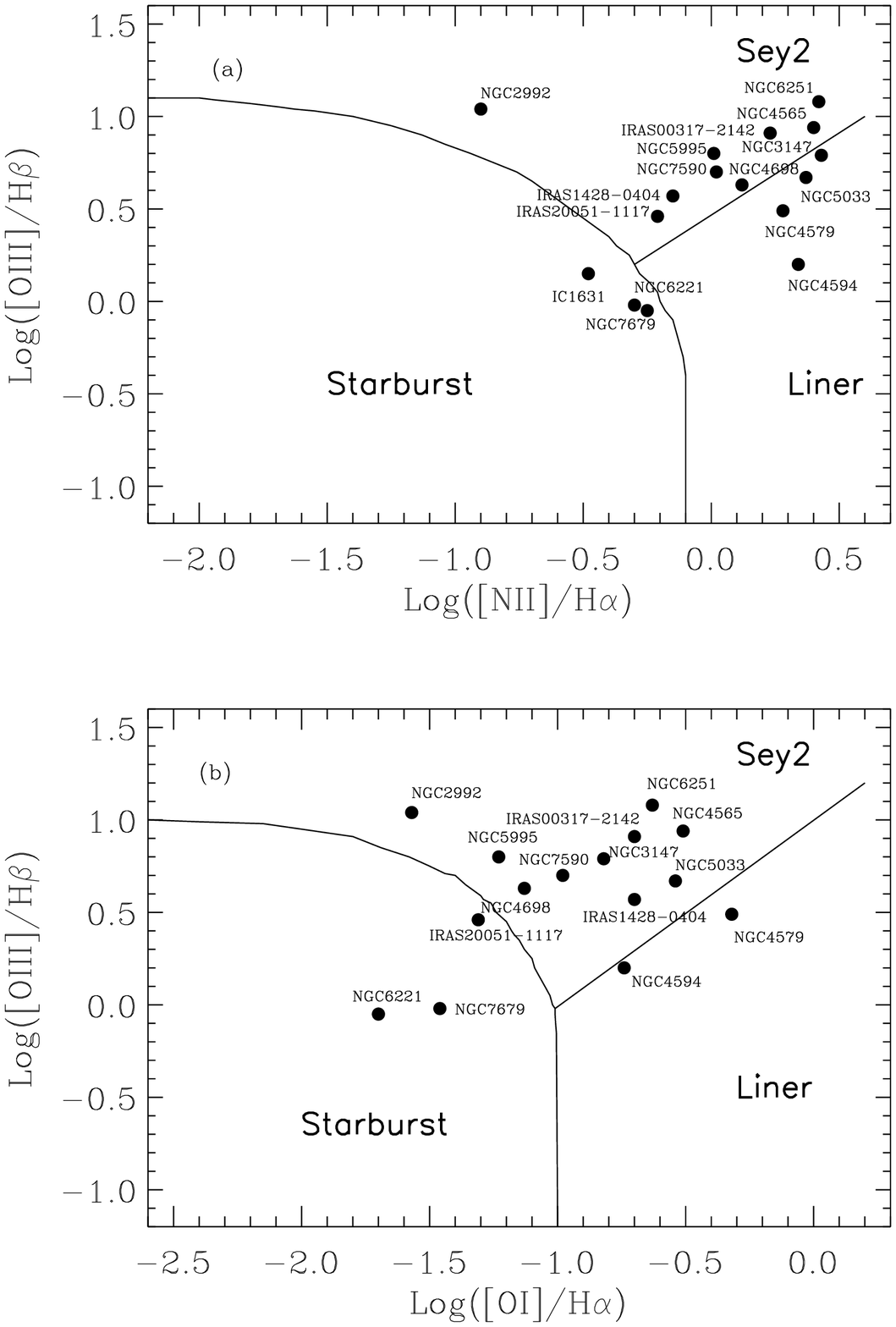,height=14cm,width=9cm,angle=0}
\caption{(a) Diagnostic diagram with [OIII]$\lambda$5007/H$_{\beta}$ vs. 
[NII]$\lambda$6583/H$_{\alpha}$. (b) Diagnostic diagram with 
[OIII]$\lambda$5007/H$_{\beta}$ vs. [OI]$\lambda$6300/H$_{\alpha}$. In both
(a) and (b) diagrams Seyfert 2, Starburst and HII regions are divided 
by solid lines.}
\label{fig:dia1_new}
\end{center}
\end{figure} 

All objects in table \ref{tab:tab_campione} are reported as pure or composite 
Seyfert 2s in NED.
In order to check the NED Seyfert 2 classification accurately,
we have employed the two optical diagnostic diagrams by 
Veilleux \& Osterbrock (1987) which use 
line-intensity ratios that are relatively insensitive to reddening 
and are considered good excitation indicators. Line ratios such as
[OIII]$\lambda$5007/H$_{\beta}$, [NII]$\lambda$6583/H$_{\alpha}$, 
[OI]$\lambda$6300/H$_{\alpha}$ delineate the different excitation mechanisms
which operate in HII regions, high-excitation AGNs (Seyferts) 
and low-excitation AGNs 
(low-ionization nuclear emission-line regions, LINERS; Heckman 1980). 

Although the boundaries between these three classes are not
rigorously defined, these diagrams represent a valid system to distinguish
between various types of narrow emission line objects.
The line ratios shown in fig. \ref{fig:dia1_new} are taken from the
literature  (see the [OIII]$\lambda$5007 flux references of table 
\ref{tab:tab_x}). For those sources with more than one observation 
we adopted the most recent reference; note that not in all cases are the 
data available or complete.

Most of the objects plotted in fig. \ref{fig:dia1_new}
show a well defined optical classification: they are classical type 2 sources.
As expected, a few objects are located at the boundaries between Seyfert 2 and 
Liner/Starburst and so are likely to be Composite objects.
The composite Seyfert/Liner nature of NGC 4579 and NGC 4594 is 
clear cut as seen in both diagrams confirming their NED 
classification; NGC 5033 and NGC 3147 are much less clear examples  as
they lie at the boundaries in one diagram but not in the other and 
so we maintain their NED definition. 

Transition objects between Seyfert 2 and Starburst are NGC 6221 and NGC 7679;
IC 1631 could be similar but unfortunately we lack information
on the [OI]/H$_{\alpha}$ ratio to confirm this hypothesis.
IRAS 20051-1117 which is classified as composite in NED is confirmed as 
this type only in one of the two diagrams and even in  this 
case it is a borderline object: we therefore take this as an indication 
of the predominance of the Seyfert 2 signature. 
Therefore we substantially confirm the NED classifications
(except in three cases which are flagged in table \ref{tab:tab_campione})
and  conclude that all  sources of our sample are characterized by an
optical type-2 signature.

\subsection{Diagnostic diagrams}
  
\input{MS2510tab2.tex}

The X-ray characteristics of the sample sources 
(described in table \ref{tab:tab_x})
strongly suggest the presence of an AGN often of 
low-luminosity in most objects (the photon indices are canonical and 
Iron lines are sometimes detected).
However this evidence is not sufficient to establish the presence of
an active nucleus in all objects and in particular in 
low luminosity sources, where the luminosity does not allow us to
discriminate between emission from an active nucleus or a starburst galaxy. 
Furthermore our objects could 
be Compton thick (N$_{H}$ $>$ 10$^{24}$ cm$^{-2}$) but since the  
photoelectric cut-off would not be detectable in the 2 - 10 keV spectrum, 
the column density  measurements would be too low. 
However the presence of an AGN and 
the Compton nature (thin or thick) of each source can 
be checked by comparing isotropic versus anisotropic properties.

If a molecular torus is present in Seyfert galaxies, then it should block the
X-ray emission coming from the central engine but it shouldn't intercept 
emission coming from larger scale structures like the Narrow Line Region
or a non nuclear starburst region. The column density could then
be inferred from the flux ratios of the X-ray fluxes versus various isotropic 
emission measurements. The [OIII]$\lambda$5007 
flux is considered a good isotropic indicator because it is produced in the  
Narrow Line Region (Maiolino \& Rieke 1995; Risaliti et al. 1999; 
Bassani et al. 1999). Also the Far-Infrared emission seems to be produced 
over a larger region than that of the molecular torus and it has been 
used as an isotropic indicator too.

The F$_{X}$/F$_{[OIII]}$ ratio has been studied in a large sample of 
Seyfert 2 galaxies: all Compton thin Seyferts show ratios higher than 
$\sim$ 1 while Compton thick sources show ratios below this value
(Bassani et al. 1999).
The  F$_{X}$/F$_{IR}$ ratio has also  been largely discussed
and used in the literature to investigate the presence of 
high column densities. 
Typically type 1 and Compton thin type 2 AGN 
show ratios of $\sim$ 0.1, while Compton 
thick type 2 objects show ratios lower than 5 $\times$ 10$^{-4}$
(David et al. 1992; Mulchaey et al. 1994; Risaliti et al. 1999).
Finally, since infrared emission is associated mainly  with 
star-forming activity while the [OIII]$\lambda$5007 emission is produced 
by photons generated in the active nucleus, by comparing the 
F$_{[OIII]}$/F$_{IR}$ ratio of all the starbursts and AGNs in the 
Ho et al. (1997) sample, Bassani et al. (in preparation) find that 
$\sim$ 90\% of the starburst galaxies
show a value below 10$^{-4}$ while $\sim$ 88\% of AGNs show a value above 
10$^{-4}$, therefore they suggest this value as a means of
discriminating between one galaxy class and the other.
In conclusion, these 3 ratios can provide an independent way to establish
which is the dominant component between AGN or starburst and 
at the same time they are a powerful tool in the detection of 
Compton thick sources when an X-ray spectral analysis is not sufficient.

\begin{figure}[htbp]
\begin{center}
\psfig{figure=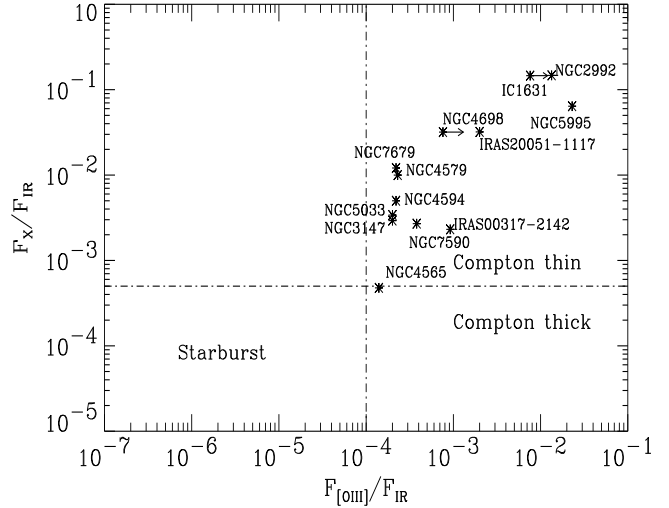,height=7.5cm,width=9cm,angle=0}
\caption{F$_{X}$/F$_{IR}$ vs. F$_{[OIII]}$/F$_{IR}$ for the sources of the
sample. We separate with dashed lines the Compton thin, 
Compton thick and Starburst region. The fluxes used are listed in
table \ref{tab:tab_x}.}
\label{fig:agno_xir}
\end{center}
\end{figure} 

\begin{figure}[htbp]
\begin{center}
\psfig{figure=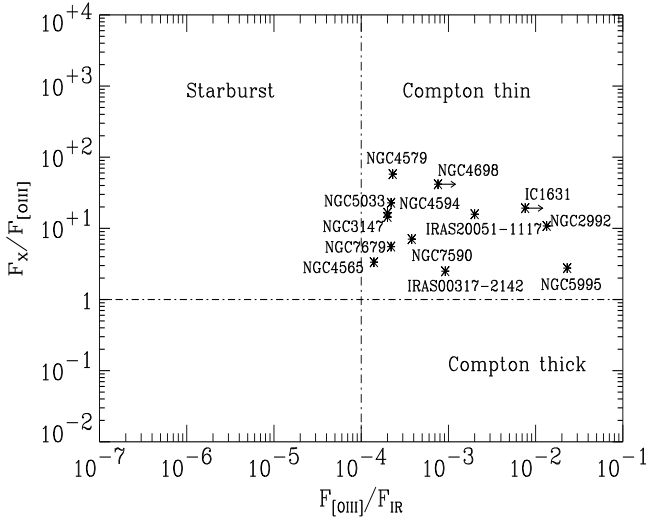,height=7.5cm,width=9cm,angle=0}
\caption{F$_{X}$/F$_{[OIII]}$ vs. F$_{[OIII]}$/F$_{IR}$ for the sources of the
sample. We separate with dashed lines the Compton thin, 
Compton thick and starburst region. The fluxes used are listed in
table \ref{tab:tab_x}.}
\label{fig:agno}
\end{center}
\end{figure} 

The fluxes used in our diagrams (figures \ref{fig:agno_xir} and 
\ref{fig:agno}) are listed in table \ref{tab:tab_x}.
All far-infrared fluxes are based on IRAS data, taken from NED. For this paper
we adopt the same definition used in Mulchaey et al. (1994) for the 
 far-infrared flux:
\begin{equation} 
       F_{IR}=F_{25\mu m} \times \nu_{25\mu m} + F_{60\mu m} \times \nu_{60\mu m}
\end{equation}  
Note that due to the IRAS angular resolution the F$_{IR}$ value
is likely to be overestimated for the nuclear region of each galaxy;
this translates into F$_{X}$/F$_{IR}$ and F$_{[OIII]}$/F$_{IR}$ ratios
smaller than in reality.
The [OIII]$\lambda$5007 flux of each galaxy has been corrected for extinction
using the formula given in Bassani et al. (1999).

In both figures \ref{fig:agno_xir} and \ref{fig:agno}
the starburst region is empty suggesting that the dominant component is 
likely to be an AGN; the region populated by our sources is that of
the Compton thin regime, indicating that indeed we measure the real 
amount of absorption in all our objects. This result is confirmed also for 
those objects for which the data are not complete: the F$_{X}$/F$_{[OIII]}$
ratios for MRK 273x, NGC 6221 and NGC 6251 are 71, 654 and 24 respectively and
the F$_{X}$/F$_{IR}$ ratio for IRAS 01428-0404 is 5.4 $\times$ 10$^{-3}$.

The major problem when dealing with low luminosity AGN is the possible 
contamination from off nuclear sources in the observed galaxy: 
in this case the measured flux is overestimated by non-imaging/low-angular 
resolution instruments and the spectrum alterated. This is what is emerging
from recent works based on {\it Chandra} and XMM-{\it Newton} observations: 
the X-ray fluxes are in most cases lower than those measured by past satellites
(see {\it Chandra} results by Ho et al. 2001). 
A consequence of this could be a mistaken evaluation of the column
density and/or of the Compton nature of the source. However, for those 
5 sources of our sample which have {\it Chandra} or XMM-{\it Newton} data the
spectral parameters are substantially in agreement with the 
old values and even if the 2 - 10 keV luminosities are somehow decreased, 
all these objects are still in the Compton thin region. 

In any case, it is important to consider the possibility that more accurate
and higher spatial resolution observations could reduce the flux
of some of our sources. 

\section{How many unabsorbed Seyfert 2 galaxies are there?} 

Bassani et al. (1999) presented a column density
distribution for a sample of Seyfert 2s for which X-ray data were available.
The distribution was shifted toward high N$_{H}$ values and only 
$\sim$ 20\% of the sources had a column density less than 10$^{22}$ cm$^{-2}$.
However, the 
sample used by Bassani et al. (1999) was a collection of data and so it 
did not fulfill the requirements of completeness. When a complete sample 
was considered (Risaliti et al. 1999) the column density distribution found 
was similar for high column density values: 75\% of type 2 Seyferts were 
found to be heavily obscured 
(N$_{H}$ $>$ 10$^{23}$ cm$^{-2}$) with 50\% of the sources being Compton thick 
(N$_{H}$ $>$ 10$^{24}$ cm$^{-2}$). However, only $\sim$ 4\% of the sample was 
characterized by low absorption (N$_{H}$ $<$ 10$^{22}$ cm$^{-2}$). 
The sample considered by Risaliti et al. (1999) was derived by 
the Maiolino \& Rieke (1995) sample completed with 
NGC 1808 and integrated with 18 new sources found by 
Ho et al. (1997) which would have been included in 
the Maiolino \& Rieke (1995) sample if they had 
been discovered earlier. This sample has been limited in
[OIII] flux (F$_{[OIII]}$ $>$ 40 $\times$ 10$^{-14}$ erg cm$^{-2}$ s$^{-1}$) 
in order to be independent of absorption effects and to have 
the greatest possible range in column density. 

We have reconsidered the total sample reported in Risaliti et al. (1999),
considering also the sources with  
F$_{[OIII]}$ $<$ 40 $\times$ 10$^{-14}$ erg cm$^{-2}$ s$^{-1}$
and we have updated the column density distribution with N$_{H}$ measurements
which  have become available in the meantime.
Finally, we have used the X-ray versus [OIII] flux ratios to assess the 
possible Compton thick nature of the sample sources, if a column density was
not available: sources found to be Compton thick were ascribed to 
N$_{H}$ $>$ 10$^{24}$ cm$^{-2}$ range. This sample consists of 92
sources and for 59 sources we have the value of the column density.
We find that the fraction of objects with N$_{H}$ $<$ 10$^{22}$ cm$^{-2}$ 
increases to $\sim$ 12\% and this is to be considered a 
lower limit as not all sources have the column density measured. 
If we consider only the 59 sources with known N$_{H}$, 
the fraction is $\sim$ 18\%.

\input{MS2510tab3.tex}

In order to check this result 
we also considered all sources which are 
classified as type 2 Seyferts (including the transition objects indicated 
as Liner2/Seyfert2 or Transition2/Seyfert2) in the Ho et al. (1997) 
sample, which is complete in magnitude to B$_{T}$ $<$ 12.0 mag. 
The total number of these objects is 49 and the column density has been
determined for 19 of them. For all these sources
the lower limit for the fraction of objects 
with N$_{H}$ $<$ 10$^{22}$ cm$^{-2}$ 
still remains around 10\%, while for only those with known N$_{H}$ (19) 
the fraction is $\sim$ 26\%. If we limit the sample in distance to
$<$ 22 Mpc, in order to limit the number of sources without a column density
measurement, we obtain a sample of 27 sources and 14 objects have 
an estimation of the N$_{H}$. In this case, the lower limit for the 
fraction increases to 15\%, while for the 14 sources this fraction is 
$\sim$ 28\%. 
In table \ref{tab:tab_fra} we show the lower limits for 
the column density distributions derived in the three samples. 

Although the number of sources for which there is no information 
available is high in all samples, the agreement found is reassuring. 
It is also interesting to note that an important fraction of unabsorbed 
objects is found only when low luminosity AGNs (LLAGNs) are considered: 
this is evident in the {Risaliti et al. (1999) total} sample when 
objects at low luminosity are included and it is obvious in the 
Ho et al. (1997) sample which is mainly made up by such sources. 

Recently, Mainieri et al. (2002) have presented the results of the
X-ray spectral analysis of the deep survey of the Lockman Hole field 
with the XMM-{\it Newton} observatory. Nearly 30\% of type 2 Seyfert galaxies
show N$_{H}$ $<$ 10$^{22}$ cm$^{-2}$.

If confirmed by future studies, the fraction of low absorbed
Seyfert 2 galaxies obtained (presumably in the range of 10\%-30\%) could 
have important implications for synthesis models
of the X-ray background and the Unified Theory of Seyfert galaxies.

\begin{figure}[htbp]
\begin{center}
\psfig{figure=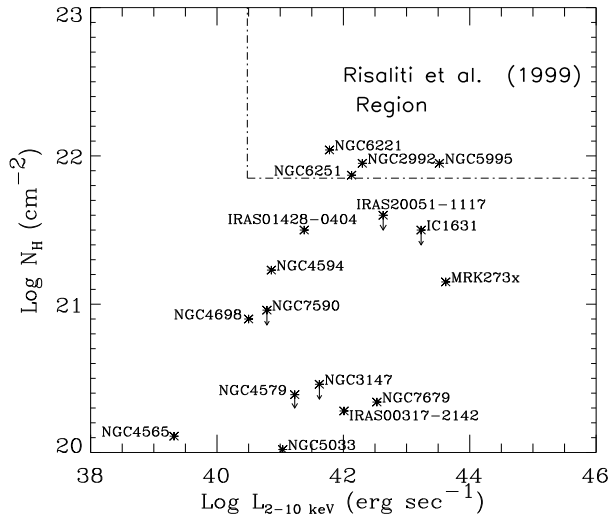,height=7.5cm,width=9cm,angle=0}
\caption{X-ray column density vs. L$_{2 - 10 keV}$ for the sources sample. 
No significant correlation between the two quantities is found. 
We highlight the region covered by Risaliti et al. (1999) sample 
sources and show that most of our objects are LLAGNs.}
\label{fig:lum1}
\end{center}
\end{figure} 

In fig.\ref{fig:lum1} we plot the column density, 
measured from the X-ray spectra, against 2 - 10 keV luminosity corrected 
for absorption. As expected we find no significant correlation
(at 90\% level for a two tailed test)
between these two quantities confirming previous 
findings by Risaliti et al. (1999).
Nevertheless it is interesting to note that we have extended the parameter
space in fig. \ref{fig:lum1} considerably towards lower N$_{H}$ and
X-ray luminosities compared to Risaliti et al. (1999) work: this
further supports the idea that more unobscured Seyfert 2s 
are observed going towards low luminosities. 
Since the LLAGNs constitute the majority
of the nearby AGN population, it becomes evident that unobscured 
type 2 objects may constitute a non-negligible part of the type 2 class. 

\section{What kind of objects are unabsorbed Seyfert 2s?} 

None of the sources in our sample show evidence of the broad 
(FWHM $>$ 1000 km/s) permitted emission lines observed in Seyfert 1s. 
However, some of our sources show a broad H$_{\alpha}$ component 
(NGC 2992, Gilli et al. 2000; NGC 4579, Ho et al. 1997; 
NGC 4594, Faber et al. 1997; NGC 5995, Lumsden \& Alexander 2001; 
NGC 7679, Kewley et al. 2000); this is sometime variable 
(NGC 5033, Ho et al. 1995) and sometime extremely 
weak (IRAS 00317-2142, Coziol et al. 1993; NGC 4565, Ho et al. 1997). 

Polarized spectroscopic measurements are available only for three objects: 
for both NGC 2992 (Rix et al. 1990) and NGC 5995 (Lumsden \& Alexander 2001 
and Tran 2001) a polarized broad H$_{\alpha}$ line has been measured, 
while for NGC 7590 no polarized broad H$_{\alpha}$ component has been detected 
(Heisler et al. 1997).

We have indications that at least 8 sources have broad components,
but the broad components are not necessarely associated with an
AGN Broad Line Region since they are too narrow for BLR standards.
It is likely that the unabsorbed Seyfert 2 galaxies
host a hidden BLR but we cannot rule out that in some of these objects
the BLR is intrinsecally weak or absent. Therefore two scenarios are plausible:
\begin{itemize}
\item A standard BLR exists, however to provide a type 2 classification
the BLR must be hidden by a non-standard absorbing medium.
\item The BLR is non-standard (very weak or fading away) or it does
not exist at all, in this case no absorbing material is required and our 
sources are truly not absorbed.  
\end{itemize}

A possible solution within the first scenario is to 
assume a gas-to-dust ratio different from the Galactic one.
This is a reasonable assumption since it has  already been 
shown to be the case for a number of Seyfert galaxies (Maiolino et al. 2001). 
However, contrary to previous studies, in our case we need a dust-to-gas 
ratio higher than Galactic: in fact galaxies with 
N$_{H}$ $\sim$ 10$^{21}$ cm$^{-2}$
would have an extinction A$_{V}$ $\sim$ 0.45 insufficient to hide the BLR
unless the A$_{V}$/N$_{H}$ is a factor 10 - 50 higher than the Galactic value.
This points to dust properties in our galaxies which are anomalous with 
respect to the great majority of AGNs. This result confirms previous findings
of Maiolino et al. (2001) who notice that a few LLAGNs in their sample were
characterized by an A$_{V}$/N$_{H}$ ratio consistent  with
or even higher than the Galactic value.

Indeed, in the HST imaging survey of the nearest Seyferts, Malkan et al. (1998)
found that the center regions of Seyfert 2 galaxies are intrinsically 
more dusty than those of Seyfert 1 galaxies, this property manifests 
itself in the form of lanes and patches that are irregular and reach 
close to the nucleus.
However, this dusty and irregular environment is likely to be located on 
larger scale than the BLR and is likely to be associated with the 
inner regions of the host galaxy (Galactic Dust Model, GDM). 

Another large scale structure which could be responsible for the absorption
measured in X-rays is a starburst region (Weaver 2001; Levenson et al. 2001a,
Starburst Model, SBM).
Such regions can have large X-ray column densities of up to a few times
10$^{22}$ cm$^{-2}$. Such values can obscure the central AGN at optical
wavelengths making a Seyfert 1 galaxy into a Seyfert 2 without requiring a 
pc-scale torus. This model also explains a higher presence of dust lanes,
bars and star-forming activity in type 2 Seyferts
than in type 1 Seyferts (Maiolino et al. 1997). Some objects of our sample are
indeed characterized by the presence of a bar (IRAS 00317-2142, IRAS 
01428-0404, NGC 4579, NGC 5995, NGC 6221 and NGC 7679) or emission from an 
HII region comparable to those of starburst galaxies (IC 1631, NGC 3147 
and IRAS 20051-1117) and can therefore be explained within the GDM and SBM. 

We can therefore check if the absorption present in our Seyfert 2 sample
is compatible with these two models. This is feasible since
in our objects the optical extinction at larger scales (i.e. in the NLR) can be
assessed from the narrow emission lines using the narrow H$_{\alpha}/H_{\beta}$
ratios and applying the Ward et al. (1987) formula:

\begin{equation}{A_{V}=6.67\times(log(H_{\alpha}/H_{\beta})-log(2.85)) \ \ \ 
(mag)}
\end{equation} 

\begin{figure}[htbp]
\begin{center}
\psfig{figure=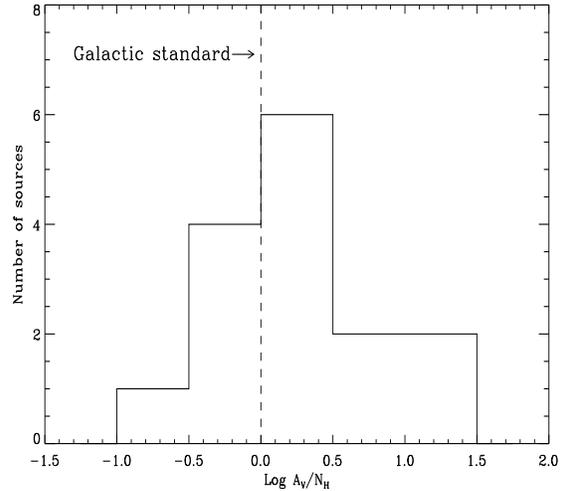,height=7cm,width=8cm,angle=0}
\caption{Log A$_{V}$/N$_{H}$ distribution for the sample. The dashed line is 
relative to the standard Galactic value.}
\label{fig:av}
\end{center}
\end{figure}

These values can then be translated into gas column densities using equation
2 from Gorenstein (1975) and compared to the N$_{H}$ values 
measured from X-ray data to obtain the A$_{V}$/N$_{H}$ ratio relative to the 
NLR. This A$_{V}$/N$_{H}$ distribution is plotted in figure \ref{fig:av} 
where it is evident that the obscuration seen in X-rays is consistent 
with that measured in the NLR, expecially after considering the large 
errors affecting both the optical and X-ray measurements.

This conversely implies that in our sources the obscuring torus is either
not visible or not present. This result conforms with both the GD and SB models
in the sense that the absorbing material in X-rays is located
at larger scale and that the A$_{V}$/N$_{H}$ ratio measured in our sources 
allow for a dust-to-gas ratio equal to or even higher than Galactic. 
In this second hypothesis the optical extinction could be sufficient to 
hide the BLR and provide the type 2 classification 
of our sources.

In the second scenario no absorption is required to hide the broad 
line region which is therefore absent or very weak. This scenario is
more likely applicable to low luminosity AGNs where the brightness
of the active nucleus may be insufficient to photoionize the BLR.

\section{Discussion}

Under the Unified Theory hypothesis, when we are looking at a Seyfert 2 we
should expect to measure a high X-ray column density produced
by a thick molecular torus which hides the BLR. According to 
the results detailed above, the column density measured in our 
Seyfert sample is consistent with 
the extinction obtained from the reddening of the lines produced in the NLR.
This means that the presence of an obscuring torus is not required and,
therefore, that the objects considered in our sample are somehow anomalous 
Seyfert 2 galaxies.
A column density of N$_{H}$ $\sim$ 10$^{22}$ cm$^{-2}$ or slightly under this 
value can be associated with the presence of a dust lane, a bar or an 
HII region. This could be the case for the objects in the present sample which
in figure \ref{fig:lum1} populate the high X-ray luminosity - high column
density region (IRAS 01428-0404, IC 1631, NGC 2992, NGC 5995, NGC 6221, 
NGC 6251 and IRAS 20051-1117); these have
N$_{H}$ values sufficient to cover the BLR.  These objects could be
seen perpendicularly to the torus and some other type of absorbing
material obscures the BLR. To summarise, if the BLR 
of these galaxies is obscured, then a larger-scale dusty environment is likely
to be responsible for their obscuration.

However, in figure \ref{fig:lum1} at lower X-ray luminosities and lower 
column densities we find those objects with N$_{H}$
very low and sometimes compatible with the Galactic 
absorption; i.e. a value not
sufficient to obscure the BLR (IRAS 00317-2142, NGC 3147, NGC 4579, 
NGC 4594, NGC 4565, NGC 4698, NGC 5033, MRK 273x, NGC 7590 and NGC 7679).
A dust-to-gas ratio different from the Galactic one could be a possible
solution to the problem, i.e. a strong dust contribution obscures the BLR 
and it is not detectable in X-rays. It is worth noting that, except for 
MRK 273x and NGC 7679, these low N$_{H}$ Seyferts are all LLAGNs.
The nature of LLAGNs is a hotly debated issue, since it is not clear
what powers such objects and it is not obvious that the 
unified model is valid at these low luminosities (Ho et al. 2001). 

A few of our galaxies have very massive black holes ($>$ 10$^{9}$ M$_{\odot}$
for NGC 4594 and 6 - 13 $\times$ 10$^{7}$ M$_{\odot}$ for NGC 4579) and they
accrete at extremely sub-Eddington rates (L/L$_{Edd}$ $<$ 10$^{-5}$, Ho 1999).
It is possible that in these extreme conditions there is not sufficient 
ionization power to illuminate the BLR or there is insufficient gas 
to feed the clouds in this region or the structure of the accretion flow is
uncapable of activating it (Barth 2002). 
Such objects would be preferentially detected in early type galaxies 
with massive black holes because a sub-Eddington accreting AGN 
would be too faint to be visible in a spiral with a low black hole mass. 
About 50\% of our LLAGNs are in early type galaxies.
There is observational evidence to support the above suggestion: 
several recent works have investigated the size-mass-luminosity relations 
in AGNs (Wandel et al. 1999; Kaspi 2000; Collin \& Hur\'{e} 2001)
supporting the idea that there is a correlation between the luminosity of the 
AGN and its BLR size in the sense that the BLR size decreases as the AGN 
luminosity decreases.

There are also theoretical studies which lead us to believe that in 
intrinsically weak AGNs the BLR is extremely faint or absent.
Nicastro (2000) suggests a model in which the FWHMs of the broad
emission lines are related to the accretion rate\footnote{In this model the 
broad emission lines originate in a vertical disk wind, at a 
critical distance in the accretion disk and the widths of such broad 
emission lines are the Keplerian velocities of the accretion disk at the
radius at which the wind originates. The disk wind forms for external accretion rates higher than a minimum value below which a standard disk is stable.}, 
in particular at very low accretion rates the clouds of the BLR would 
cease to exist. Therefore for sufficiently weak AGNs no BLR is expected.
This is the 'pure' Seyfert 2 model (Tran 2001).

Furthermore, we consider the hypothesis of 'Fossil' objects  
in which the central engine has turned off (Maiolino 2000). Shortly after this 
occurs, the BLR fades away and the nucleus appears as a type 2 object.
If the nucleus remains quiescent for a sufficient time then 
even the echo of the NLR will fade away. Under this hypothesis 
unabsorbed Seyfert 2s are simply a transition state of an AGN. Since this phase
should rapidly decay, the fraction of fossil AGNs may be very small.  

The obscuring material around an active nucleus has been proven by
some recent evidence to be complex and variable and the possibility that 
in some objects our line-of-sight might be intercepting a 'hole' 
in the absorber at a certain time must also be taken into account 
(Risaliti 2002).

To summarize, in the 17 unabsorbed Seyfert 2 galaxies 
the low absorption is not likely to be related to the 
torus but to material distributed over larger scale, 
very likely the NLR unless the BLR is particularly dusty. 
Some of these objects could be seen perpendicularly to the torus with 
some types of absorbing material obscuring the BLR; the 
material may be slightly dusty and associated with lanes, patches or 
starburst regions. In a few cases, the lack of evidence for the existance 
of a BLR suggests that this is weak, absent or has just faded away. 
In this case the nucleus is of low luminosity 
and this may explain the weakness or lack of a BLR.

\section{Conclusions}

We present a sample of type 2 Seyfert galaxies which behave differently than
classical type 2 objects. We have shown that there are strong indications
in favour of the 'true' Compton thin nature of such sources.

We estimate the fraction of Seyfert 2s with low absorption to be
in the range of 10\%-30\% and this fraction is likely to increase 
at low luminosities. The study of a complete sample of LLAGNs 
will provide a more reliable estimate of their incidence in the 
Seyfert galaxies population.

The unabsorbed Seyfert 2 galaxies cannot be described by the Unified 
Model in its simple formulation. The standard BLR of Seyfert galaxies 
is either obscured by something other than the torus or it is weak, 
absent or has just faded away.

As a concluding remark, it is worth noting that the results presented in this 
paper add further complications to the classification scheme of AGNs in general
and of Seyfert galaxies in particular.

\begin{acknowledgements}

We thank Massimo Cappi and Luigi Foschini for useful comments
and John Stephen for the help in the final revision. We thank the referee
for the helpful comments that improved the manuscript. 
This work has made use of XMM-{\it Newton} data.
This research has made use of the NASA/IPAC extragalactic 
database (NED) which is operated by the Jet Propulsion Laboratory, 
California Institute of Technology, the High Energy Astrophysics 
Science Archive Research Center (HEASARC) provided by NASA's Goddard Space 
Flight Center, the {\it Beppo}SAX Science Data Center, the ASCA archive and
the TARTARUS database.

\end{acknowledgements}

\end{document}

%% file: MS2510tab1.tex
\renewcommand{\tabcolsep}{3.9pt}

\begin{table}[!htbp]
\bigskip
\caption{\bf Unobscured Seyfert 2 galaxies sample}
\label{tab_camp}
\vspace{0.5cm}
\scriptsize
\label{tab:tab_campione}
\begin{center}
\begin{tabular}{l l c c c}
\hline\hline
\\
Nome &RA \&Dec. (2000) & $z$ & N$_{\it HGal}^{\S}$ & NED\\
      &                 &  & & class.\\
\\
\hline
\\
IRAS 00317$-$2142 &00 34 13.8  $-$21 26 21  & 0.02680 & 1.5  & S1.8    \\
\\
IRAS 01428$-$0404 &00 45 25.2  $-$03 49 36  &0.01820 & 4.28  & S2 \\
\\
IC 1631         &01 08 44.8  $-$46 28 33  &0.03084  &2.17    & S2$^\star$ \\
\\
NGC 2992        &09 45 42.0  $-$14 19 35  & 0.00771  & 5.26  & S2 \\
\\  
NGC 3147        &10 16 53.6  +73 24 03  & 0.00941  & 3.64    & S2 \\
\\
NGC 4565        &12 36 20.8  +25 59 16  & 0.00428  & 1.30    & S1.9 \\
\\
NGC 4579        &12 37 43.5  +11 49 05  & 0.00507 &2.47      & L/S1.9  \\
\\
NGC 4594        &12 39 58.8  $-$11 37 28  & 0.00364 &3.77    & L/S1.9   \\
\\
NGC 4698        &12 48 23.0  +08 29 14  & 0.00334 &  1.87    & S2    \\
\\
NGC 5033        &13 13 27.3  $+$36 35 36 &0.00292 &  1.03    & S1.9   \\
\\
MRK 273x        &13 44 47.4  $+$55 54 11  &0.45800 &1.10     & S2 \\
\\
NGC 5995        &15 48 24.9  $-$13 45 28  & 0.02519  & 10.6  & S2   \\
\\
NGC 6221        &16 52 46.1  $-$59 13 07  & 0.00494 & 15.0   & S2$^\star$ \\
\\
NGC 6251        &16 32 31.8  +82 32 16  & 0.02302 &5.49      & S2  \\
\\
IRAS 20051$-$1117&20 07 51.3  $-$11 08 33 &0.03149 &6.8      & SB/S2 \\
\\
NGC 7590        &23 18 55.0  $-$42 14 17  & 0.00532  & 1.96  & S2   \\    
\\
NGC 7679        &23 28 46.8  +03 30 41  &0.01714  &5.13      & S2$^\star$ \\
\\
\hline\hline

\end{tabular}
\end{center}
\smallskip
\S Column density in units of 10$^{20}$ cm$^{-2}$ \\
$\star$ Transition objects between Seyfert 2 and starburst (see figure 
\ref{fig:dia1_new}) \\
\end{table}

%% file: MS2510tab2.tex
\renewcommand{\tabcolsep}{8.8pt}

\begin{table*}
\bigskip
\caption[]{\bf X-ray spectral parameters for the sample}
\begin{center}
\scriptsize

\label{tab:tab_x}
\begin{tabular}{llcc ccc  cc}
 \\ \hline\hline
\\
 & Nome & $\Gamma$ & N$_{\it Hint}^{a}$ & EW(Fe$_{k\alpha}^b$) & 
F$_{2 - 10 keV}^c$ & L$_{2 - 10 keV}^d$  & F$_{IR}$$^e$ 
& F$_{[OIII]}$$^f$  \\
\\
\hline
\\
1&IRAS 00317$-$2142   & 2.00$^{+0.07}_{-0.07}$ & 1.9$^{+0.2}_{-0.2}$ &
$<$900  & 0.08 &  42.01 &25.94 &24 \\
\\
2&IRAS 01428$-$0404  &1.95$^{+1.33}_{-0.73}$ & 32$^{+109}_{-32}$ & $-$&
0.04& 41.38 & $<$7.42 & $-$\\
\\
3&IC 1631     & 2.10$^{+0.10}_{-0.10}$ &      $<$ 31.6         &
 $<$ 70               & 1.00  &  43.23 & $<$6.85 & 52.0 \\
\\   
4&NGC 2992   & 1.70$^{+0.2}_{-0.2}$        &  90.0$^{+3.0}_{-3.0}$      &
147$^{+37}_{-37}$ & 7.4& 42.30 & 50.79 & 680.0\\
\\
5&NGC 3147    &   1.94$^{+0.20}_{-0.19}$  & $<$2.9 &
675$^{+395}_{-328}$       & 0.22  & 41.62  &44.39 &9.0 \\
\\
6&NGC 4565    &   1.7$^{+0.2}_{-0.2}$ & N$_{H,gal}$ &  $-$  &
0.02 &  39.32& 42.03&6.0   \\
\\
7&NGC 4579 &   1.88$^{+0.03}_{-0.03}$       & N$_{H,gal}$ &
$-$       & 0.52  &  41.23 &38.37 & 9.0 \\
\\
8&NGC 4594 &  1.5$^{+0.04}_{-0.03}$      & 17$^{+11}_{-9}$ &
 $-$ & 0.16  &  40.86 & 31.90 &7.0  \\
\\
9&NGC 4698 &1.91$^{+0.14}_{-0.14}$&8.1$^{+8.2}_{-7.8}$&$<$425&0.10& 40.50 &
$<$3.14 &2.4  \\
\\
10&NGC 5033    & 1.7$^{+0.02}_{-0.02}$ & N$_{H,gal}$ &
100$^{+100}_{-100}$ & 0.28 &  41.04 & 82.77 &17.0 \\
\\
11&MRK 273x  & 1.66$^{+0.15}_{-0.11}$ & 14.1$^{+5.5}_{-5.0}$&
$<$30& 0.01&  43.62& $-$ & 0.14\\
\\
12&NGC 5995  &1.81$^{+0.04}_{-0.03}$&90$^{+5}_{-3}$&144$^{+41}_{-41}$&2.89&
 43.52& 28.54 & 66 \\
\\
13&NGC 6221 &1.9& 110$^{+8.6}_{-8.3}$ & 360$^{+210}_{-93}$$^{g}$ & 1.4& 
 41.78& $-$ &2.14\\
\\
14&NGC 6251   &  1.83$^{+0.21}_{-0.18}$      & 75$^{+64}_{-58}$
& 443$^{+313}_{-272}$$^{g}$ & 0.14 & 41.13 &$-$ & 57.0  \\
\\
15&IRAS 20051$-$1117 &1.92$^{+0.21}_{-0.14}$ & $<$40 & 272$^{+52}_{-73}$ &
0.24&  42.63 &7.59& 15.2 \\
\\
16&NGC 7590    & 2.29$^{+0.20}_{-0.13}$        &$<$ 9.2      &
  $-$     &0.12  &  40.79 & 44.53 & 17.0 \\
\\
17&NGC 7679    & 1.75$^{+0.03}_{-0.06}$  & 2.2$^{+1.5}_{-1.4}$  &   $<$200  & 
0.60  &   42.53 & 49.60 & 108.26  \\
\\
\hline\hline
\\
\end{tabular}
\end{center}
\smallskip
(a) Column density in units of 10$^{20}$ cm$^{-2}$ (b) Equivalent width 
of the Fe$_{k\alpha}$ in units of eV (c) 2 - 10 keV flux in units 
of 10$^{-11}$ erg cm$^{-2}$ sec$^{-1}$ (d) 2 - 10 keV logarithm of 
unabsorbed luminosity (H$_{0}$ = 75 km s$^{-1}$ Mpc$^{-1}$) 
(e) Infrared flux in units of 10$^{-11}$ erg cm$^{-2}$ sec$^{-1}$
(f) [OIII]$\lambda$5007 corrected flux in units of 10$^{-14}$ erg 
cm$^{-2}$ sec$^{-1}$ (g) Line at 6.65 keV . X-ray references: 
(1) Georgantopoulos et al. 2000; 
(2) This work,  ASCA data; (3) Awaki et al. 1992; (4) Gilli et al. 2000; 
(5) This work,  SAX data; 
(6) Cappi et al. 2002; (7) Eracleous et al. 2001;
(8) Pellegrini et al. 2002; (9) Pappa et al. 2001; 
(10) Cappi et al. 2002; 
(11) Xia et al. 2002; (12) This work,  ASCA data; 
(13) Levenson et al. 2001b; 
(14) Sambruna et al. 1999; (15) This work,  ASCA data; 
(16) Bassani et al. 1999; (17) Della Ceca et al. 2001. All F$_{IR}$ are based
on IRAS data taken from NED. [OIII] references: (1) Moran et al. 1996; (2)
Pietsch et al. 1998; (3) Sekiguchi et al. 1993; (4) Gilli et al. 2000;
(5) Ho et al. 1997; (6) Ho et al. 1997; 
(7) Ho et al. 1997; (8) Ho et al. 1997; (9) Ho et al. 1997; 
(10) Ho et al. 1997; (11) Xia et al. 1999; (12) Lumsden et al. 2001; (13)
Levenson et al. 2001b; (14) Shuder \& Osterbrock 1981; (15) Moran et al. 1996; 
(16) Vaceli et al. 1997; (17) Kewley et al. 2000.\\ 
\\
\end{table*}

%% file: MS2510tab3.tex
\renewcommand{\tabcolsep}{8.8pt}
\begin{table}
\bigskip
\caption[]{\bf Fractions for three samples}
\scriptsize
\label{tab:tab_fra}
\begin{tabular}{l c c c}
 \\ \hline\hline
\\ 
log(N$_{H}$) & Risaliti(99) & Ho et al. (97) & Ho sample \\
cm$^{-2}$   &  total sample & sample & $<$22 Mpc \\
\\
\hline
\\
$<$22& $>$12\% & $>$10\% & $>$15\% \\
22-23& $>$14\% & $>$ 4\% & $>$ 7\% \\
23-24& $>$16\% & $>$14\% & $>$19\% \\
$>$24& $>$22\% & $>$10\% & $>$11\% \\
no data & 36\% & 62\% & 48\% \\
\\
\hline\hline
\\
\end{tabular}
\smallskip
\\
References for N$_{H}$: Ho (1999), Allen et al. (2000), 
Terashima et al. (2000), Guainazzi (2001), Pappa et al. (2001), 
Levenson et al. (2001b), Cappi et al. (2002). 
\\
\end{table}